# Unconventional light scattering from glassy photonic films and metasurfaces

Artem D. Sinelnik[1*], Kirill B. Samusev[1,2], Mikhail V. Rybin[1,2,*], and Mikhail F. Limonov[1,2]

[1] ITMO University, St. Petersburg 197101, Russia
[2] Ioffe Institute, St. Petersburg 194021, Russia
*Corresponding Author: E-mail: artem.sinelnik@metalab.ifmo.ru

Abstract

The propagation of light through a random medium is an important problem in photonics. When the random fluctuations of the orientation for individual rods were introduced to the ideal woodpile photonic structure, a crossover from Laue diffraction to randomly scattered fields which is similar in appearance to speckle patterns was observed and investigated. Unexpected interplay between order and disorder was discovered from anisotropic glassy samples when orientational disorder was added only in one direction of square woodpile structure. It is found that the ordered sets of rods produced disordered patterns and vice versa the disordered sets of rods produced ordered patterns that continue to be bright and sharp with increasing disorder. To explain this effect, it was demonstrated theoretically and experimentally that the light scattering can be described purely in terms of the intersection points of the rods.

## I. INTRODUCTION

The interplay of order and disorder in photonic structures is of fundamental importance to the propagation of electromagnetic waves and is the key mechanism for the localization and trapping of light in dielectric materials[1,2]. Whereas ideal photonic crystals take advantage of the periodicity in the dielectric contrast and the consequent long-range correlation, random media can dramatically affect all wave processes. Examples of strongly modified light transport in disordered structures include weak[3,4] and strong Anderson localization[5] of electromagnetic radiation, Fano resonance in disordered dielectric structures[6], random lasing[7], speckle correlations[8]. Random media can be used to generate a sharp focus[9], or multiple foci[10], allows the manipulation of the temporal response of the system[11]. Wavefront shaping with disorder-engineered metasurfaces allows obtaining high resolution and a large field of view at the same time, well beyond what is possible with ordered structures[12].

An ideal 3D photonic crystal consists of periodic arrangements of building blocks which are identical to each other with size comparable to the wavelength of light. Alternatively, one can also consider an interesting case of an irregular media that is a completely random arrangement of identical building blocks, which by analogy can be termed 3D photonic glasses[13,14]. The most important property of both solid systems is the monodispersity of the building blocks that compose them, for example monodisperse spheres. The broken symmetry of the photonic glasses leads to very different electromagnetic properties in comparison with photonic crystals. A laser light diffraction experiment demonstrates a Bragg diffraction pattern for the photonic crystal. For the photonic glasses a hallmark is a speckle pattern in diffraction experiments[14]. Laser speckle[15] is an interference pattern produced by light scattered from different parts of the illuminated disordered object and can only be described statistically[16]. The intensity at any point on the image on the screen is determined by the algebraic addition of all the wave amplitudes arriving at the point.

In this study, we take a new step forward in the preparation of disordered media and introduce a new type of photonic structure: glassy metasurfaces. A difference between 2D photonic structures named as metasurfaces[17, 18] and introduced here glassy metasurfaces is identical to the difference between 3D photonic crystals and 3D photonic glasses: the first medium is ordered and the second medium has extremely disordered structure. Contrary to the well studied 3D photonic glasses consisting of monodisperse spheres[13,14], we create a set of ordered and disordered woodpile structures[19] composed of several layers (from 2 to 10 layers) build up from monodisperse square rods with random fluctuations in the orientation within woodpile layer. The structures consisting of two layers only (one layer ordered and another one disordered) we consider as anisotropic glassy metasurfaces.

## II. SAMPLES PREPARATION

The problem of fabrication of true 3D ordered and disordered photonic structures of almost arbitrary shape can be solved with the method of direct laser writing (DLW)[20-23].

The aim of this work was the synthesis, structural and diffraction studies of ordered and disordered woodpile photonic structures. As this was done in our previous studies[24,25], to realize the DLW technique we use the installation and software package from Laser Zentrum Hannover (Germany). The structures were created using a hybrid organic-inorganic material based on zirconium propoxide with an Irgacure 369 photo-initiator (Ciba Specialty Chemicals Inc., Basel, Switzerland) with the refractive index $n = 1.52$.

The structure of the ordered woodpile is shown schematically in Figure 1a. This structure is built $xy$-layer by $xy$-layer, four layers make up the lattice period $c$ along the $z$ axis. The building blocks that compose the woodpile structures is a square rod with the height equal to $c/4$. In the first layer, the rods are parallel to each other with $a$ period along the $x$ axis, in the second layer the same rods with the same $a$ period are parallel to each other along the $y$ axis. In the general case, the woodpile structure has a body-centered tetragonal lattice, but for two parameter ratios the symmetry of the structure becomes cubic. When $c = a$, the lattice of the woodpile is body-centered cubic (bcc), and when $c = \sqrt{2}a$, the lattice is face-centered cubic (fcc).

With the same building block – a square dielectric rod – two types of photonic structures with lattice constant comparable to the wavelength of light were created: a perfectly ordered arrangement Figure 1a and a random arrangement of rods for glassy metasurfaces (two $xy$-layers) and woodpile glassy thin slabs (the number of $xy$-layers up to 10), Figure 2b-d. We fabricated the disordered woodpile structures as follows. Each individual rod in the layer (with index $i$ ) was turn about its center (along *x*- or *y*-axis) within the *xy*-layer by random angle $\alpha_i$ with respect to the ordered state ($\alpha_i = 0$ for the rods is the ordered structure). The technological process was done by employing two different kinds of random fluctuations. We employed normal and uniform distribution functions: $\sigma = p\frac{\pi}{4}, 0 \leq p \leq 1$ is the dispersion of $\alpha$ for the normal distribution and $-\alpha_{max} \leq \alpha_i \leq \alpha_{max}$ where $\alpha_{max} = p\frac{\pi}{4}, 0 \leq p \leq 1$ for the uniform distribution. Thus, the structures obtained were characterized by the parameters of the ordered woodpile *a* and *c* as well as by an additional parameter *p* characterizing the degree of disorder. All structures have external size in the *xy* plane of $50 \times 50 \, \mu m$, the lattice parameters varied in different samples in the range of $0.5 \, \mu m \leq a \leq 2.0 \, \mu m$, the number of layers along the *z*-axis *N* was ranging from two layers (a metasurface) to 10 layers (ordered or disordered 3D structures). The examples of the images of different structures obtained by a scanning electron microscope are presented in Figure 2.

## III. LAUE DIFFRACTION FROM 2D ORDERED PHOTONIC STRUCTURES

For the analysis of optical diffraction patterns from low-contrast photonic structures, it is sufficient to use the Born approximation when the diffraction intensity is determined by a product of the squares of the structure factor *S*(q), which is associated with the lattice periodicity, the scattering form factor *F*(q), which takes into account the contribution from an unit cell, and a polarization factor[26]. A comparison of the theoretical and experimental data demonstrates[24] that in our low contrast case it is sufficient to consider only the structure factor *S*(q). For the 2D metasurface with the square lattice symmetry, the position of each scatterer is determined by the 2D vector $\mathbf{r}_i = \mathbf{a}_1 n_1 + \mathbf{a}_2 n_2$, where $a_1$ and $a_2$ stand for the basis mutually perpendicular vectors ($a_1 \cdot a_2 = 0$) of the square ($a_1 = a_2 = a$) lattice. To analyze the diffraction patterns, we consider the scattering from one-dimensional linear chain of scatterers first. The conditions for the appearance of the Laue diffraction maxima in the case of the linear chain and normal incidence are described by the simple formula:

$$\theta_s = \cos^{-1}(n\lambda/a) \qquad (1)$$

where $a = |\mathbf{a}_1|$, $\lambda$ is the wavelength of incident light, and $\theta_s$ is the angle of light scattering on the chain between vectors $a_1$ and the wave vector of the scattered waves $\mathbf{k}_s$. This equation defines the diffraction selection rules in relation to the ratio between $\lambda$ and $a$ because the inverse cosine function is only defined in the interval from −1 to 1.

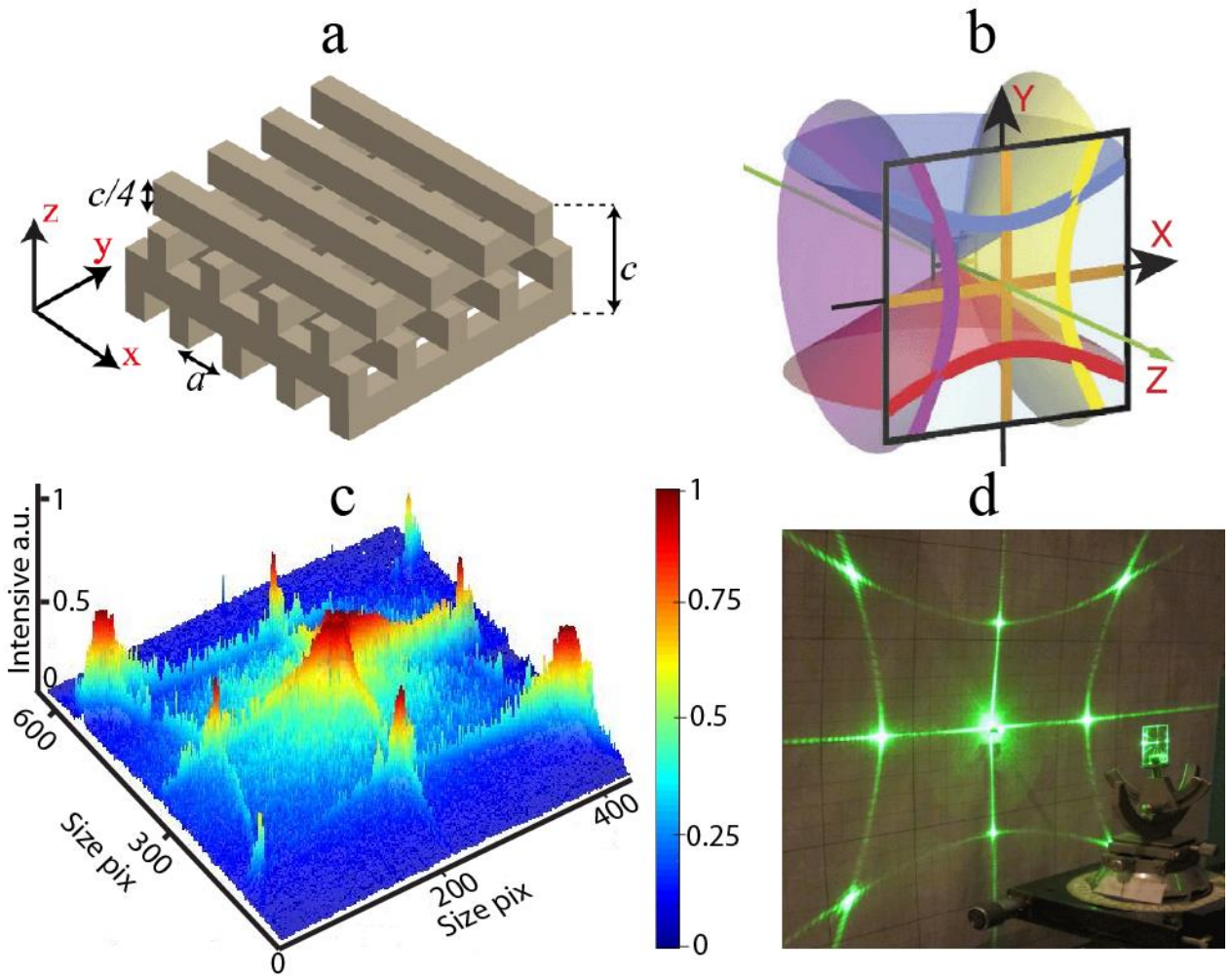

FIG. 1. a) Sketch of the woodpile structure. b) Schematic of the zero-order (n = 0) and first-order (n = ± 1) Laue diffraction from the horizontally and vertically oriented chains of scatterers for the structure with square symmetry in the case $a_1 = a_2$. Diffraction patterns on a flat screen are shown by thick lines. Scattered light is shown by different colors for clarity. c) The 3D – image of the intensity of the diffraction patterns on a flat screen placed behind the sample. The *x, y* scales are in pixels. d) Photograph of experimental setup. The zero- and first-order Laue diffraction from the ordered woodpile slab.

The zero-order diffraction ($n=0$) is observed for any ratio between $\lambda$ and *a* in the plane perpendicular to the axis **a** since the angle of light scattering becomes $\theta_s = 90°$. A pair of diffraction cones of the *n*-th order appears at $a > n\lambda$ when the structure cannot be considered as a metasurface. The ordered woodpile structure can be considered as a structure composed of two sets of mutually orthogonal chains along the x- and y-axes. Therefore the diffraction pattern consisting of zero-order scattering only (two mutually orthogonal planes) is a distinguishing feature of a metasurface ($a < \lambda$).

The experimental setup is shown in Figure 1d. The laser beam was focused by a lens (25 cm focal length) onto the sample at normal incidence. The laser spot at the samples substrate surface is about 150 μm in diameter. The diffraction patterns were observed on a flat semitransparent far-field screen placed behind the sample. The distance from the sample to the screen was nearly 20 cm. The samples are illuminated by a Nd laser with $\lambda = 0.53\,\mu m$. For all ordered structures with the square symmetry the diffraction patterns demonstrate the $C_{4v}$ symmetry, Figure 1d. We observed strong intensity of optical diffraction from rather small ($50 \times 50\,\mu m$) low-contrast dielectric samples. For different samples we can distinguish on the screen two types of the diffraction features: two orthogonal strips that correspond to the zero-order scattering ($n=0$) and pairs of arcs that correspond to the higher orders of scattering ($n= \pm 1$, $\pm 2, \pm 3...$) being formed by intersections of planes or cones with flat screen, as shown schematically in Figure 1b for $n=0, \pm 1$. For the chain of scatterers with the period of $a_1 = 1\,\mu m$ and $\lambda = 0.53\,\mu m$ the first-order cones have angle of scattering $\theta_{s1} = 58°$.

## IV. TRANSITION FROM LAUE DIFFRACTION TO SPECKLE-LIKE PATTERNS

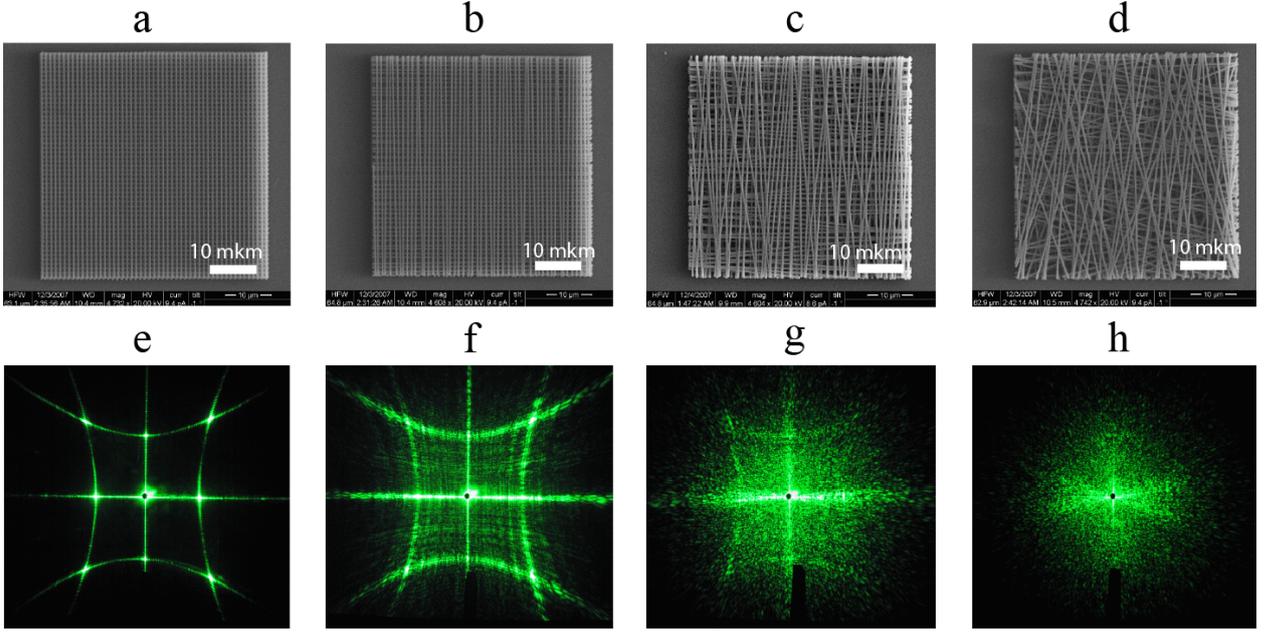

FIG. 2. The SEM image of the ordered woodpile slab, *a)* $p=0$, and glassy woodpile structures, *b)* $p=0.04$, *c)* $p=0.1$, *d)* $p=0.5$. Diffraction pattern evolution for ordered (e) and disordered (f-h) woodpile thin slabs corresponding to the samples above with normal distribution of the disordered parameter $p$. All structures have external size in the *xy* plane 1dof $50 \times 50\,\mu m$, the lattice parameters of the ordered sample $a_x = a_y = 1\mu m$, the number of layers along the *z*-axis *N*=8. The patterns are observed on a flat screen positioned behind the sample, Figure 1d. $\lambda = 0.53\mu m$.

The results of our experimental studies of light scattering from ordered and disordered woodpile thin slabs depending on the disorder parameter *p* under the normal laser incidence are presented in Figure 2 (e-h), together with the SEM imagines of the corresponding samples, Figure 2 (a-d). For ordered C$_{4v}$ square structure (Figure 2(a), $a_1 = a_2 = 1\mu m$, $\lambda = 0.53$ μm, $\lambda < a < 2\lambda$), on a far-field screen placed behind the sample one can distinguish the perfect diffraction pattern with the $C_{4v}$ square symmetry consisting of two orthogonal strips that correspond to the zero-order scattering and four arcs that correspond to the first-order scattering.

When we introduce the random fluctuations of the orientation for both *x*-oriented and *y*-oriented individual rods within the woodpiles *xy*-layer, the diffraction patterns on the screen changed dramatically Figure 2(f-h). With increasing of the disordered parameter *p*, we can distinguish two different phenomena: stripes and arcs become more and more randomized and a granular distribution of light intensity appears through the entire screen.

In our experiments, the laser spot at the samples substrate surface is about 150 *μm* in diameter and all structures have the of $50 \times 50 \, \mu m$. Therefore light from all points within the woodpile structure contributes to the diffraction intensity at any point on the far-field screen. For disordered woodpile, light from different parts of the structure traverses different optical path lengths to reach the screen. As a result, we visualize a laser speckle-like patterns which are a random interference effect that gives a high-contrast granular distribution of light intensity on a far-field screen. A strong intensity maximum is observed if all the waves arrive at the point on the screen in phase. For a sample with strong disorder $p = 0.5$ we can see a speckle-like diffraction pattern without any traces of arcs (Figure 2(h)).

It is interesting to consider the spatial evolution of the zero-order and first-order scattering more detailed. For strong disorder (*p*=0.5) there is no traces of first-order arcs however we can easily track the zero-order scattering that transforms from two orthogonal strips (*p*=0) to the pattern reminding a propeller (*p*=0.1, 0.5). The reason of the different evolution of the zero-order and first-order diffraction patterns is following. According to Equation 1, the zero-order patterns (*n*=0) do not depend on the distances between scatterers and therefore diffraction stripes we can observable on the screen from any disordered chain of scatterers. The only new feature is additional angular variation of the diffraction stripes according to the corresponding angular variation of chains of scatterers. Being always perpendicular to each individual chain of scatterers, the diffraction pattern transforms from the perfect cross to the propeller-like feature repeating the propeller-like distribution of the disordered rods. Contrary to the zero-order scattering, the first-order diffraction patterns strongly depend on the interscatterers distance accordingly to the equation $\theta_{s1} = \cos^{-1}(\lambda/a)$. Therefore for the randomized parameter $a$, the angle $\theta_{s1}$ becomes uncertain, the patterns becomes blurry and finally transform to the granular distribution of light intensity.

## V. INTERPLAY OF ORDER AND DISORDER IN OPTICAL DIFFRACTION FROM ANISOTROPIC GLASSY WOODPILE STRUCTURES

In order to gain a more complete understanding of light scattering effects on both ordered and disordered dielectric photonic structures, it is necessary to conduct a diffraction study on anisotropic glassy woodpiles. Figure 3 illustrates the evolution of the diffraction pattern from anisotropic thin slab ($N= 8$) depending on the disordered parameter $p$. The anisotropic structures are specifically designed with disorder in rods which initially oriented along $x$-axis while all rods oriented along $y$-axis are ordered (parallel to each other) in all the samples. The initial ordered structure has lattice parameters of $a_1 = a_2 = 2\,\mu m$ which makes it possible to observe both zero- and high-orders scattering with excitation wavelength of $\lambda = 0.53\ \mu m$.

It can be seen that for small amounts of disorder ($p \leq 0.01$) there is very little modification of the scattering relative to that for the ideal structure. However, with larger amounts of disorder ($p \geq 0.04$) the modification of the diffraction pattern is much more pronounced and completely unexpected. Note that $y$-oriented ordered set of rods produces horizontal stripe with vertically oriented $n$-th order diffraction cones whereas $x$-oriented disordered set of rods produces vertical stripe with horizontally oriented cones. On the screen oriented perpendicular to the sample, we can see the cones from $x$-disordered rods and the horizontal stripe with the arcs from $y$- ordered rods, Figures 3e-h. The surprising result is that with increasing of the disorder, the arcs from ordered set of rods becomes more and more randomized and finally hardly observed in Figures 3g,h while the cones from disordered set of rods continue to be strong and bright, Figure 3g. We also prepared and measured anisotropic metasurfaces with disordered top or bottom layers (that is, we switch x and y) and the results were equivalent.

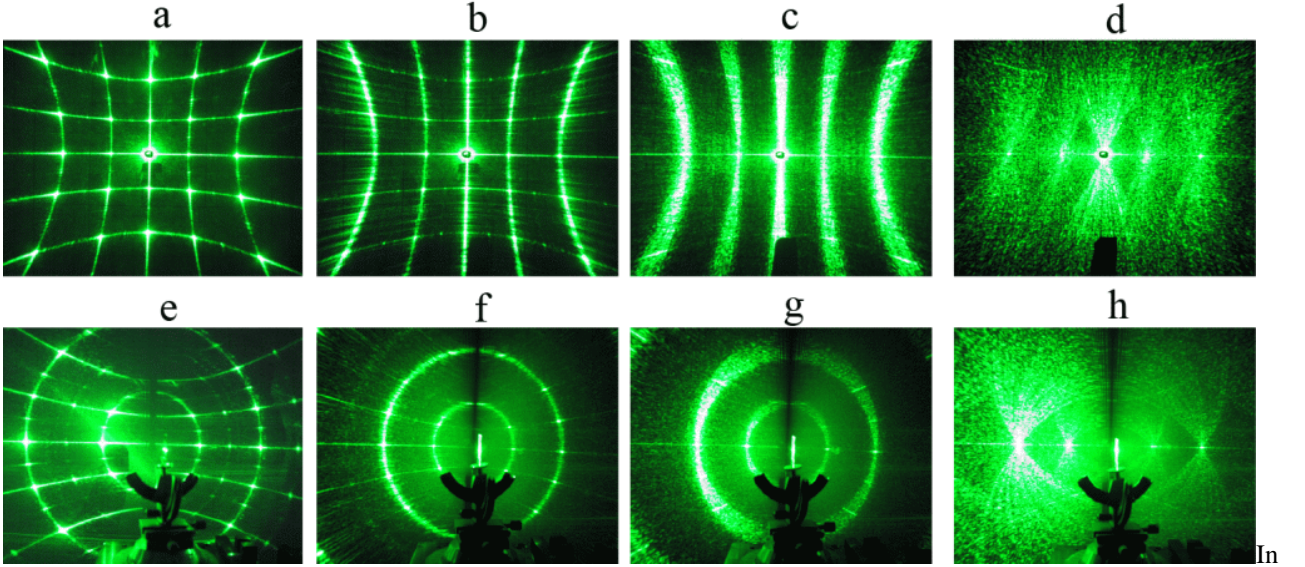

FIG. 3. Diffraction pattern evolution for anisotropic glassy woodpile thin slab depending on the disordered parameter $p$ for the normal distribution. $p = 0$ (a,e), $p = 0.01$ (b,f), $p = 0.08$ (c,g), $p = 0.5$ (d,h). All structures have external size in the $xy$ plane of $50 \times 50\ \mu m$, the lattice parameters of the ordered sample $a_x = a_y = 2\mu m$, the number of layers along the $z$-axis $N=8$. The patterns are observed on a flat screen positioned behind the sample (a-d) and perpendicular to the sample (e-h). $\lambda = 0.53 \mu m$.

At this point we face an important question: what is the nature of scatterers in the case of a structure build up of long dielectric rods? To understand the paradox of crossing linkage between structural and scattering disorder, we performed the calculations of the diffraction patterns using two different models of scatterers. We have carried out the calculations of diffraction patterns of ordered and disordered woodpile structures in the framework of the Born approximation when only the effects from a sum of single scatterings are evaluated[24]. In this approximation we assume that all scattering centers are similar and we neglect multiple scattering events. A directivity pattern of the centers is neglected too. It allows us to evaluate the scattered intensity as $I \propto \left| \sum_j \exp(i(\mathbf{k}_{inc} - \mathbf{k}_{sca})\mathbf{r}_j) \right|^2$, where $\mathbf{k}_{inc}$ and $\mathbf{k}_{sca}$ are wave vectors of incident and scattering radiation and $\mathbf{r}_j$ is a position of *j*-th scattering center. Now we have to define what are the scattering centers?

First, we have modeled each square dielectric rod as a set of 1000 coaxial equidistant point scatterers along the rod 50 μm long. Figure 4j-l show results of such calculations. Such model fails to describe the experimental disordered patters in Figure 4f because strong and sharp first-order arcs from ordered set of rods exist for high level of disorder Figure 4l. For this reason we consider another model assuming that only points of rods intersections represent light scatterers and define the scattering processes.

To explain unambiguously the paradox of crossing linkage between structural and scattering disorder, we fabricated and investigated a special type of ordered structure with two sets of rods rotated through 45° with respect to one another, Figure 4b. Such 45°-structure of roads has the symmetry of a parallelogram $C_2$ although the array of rods intersection points has the $C_{4v}$ square symmetry of the perfect woodpile structure. For clarity, the points of intersections are marked by red, blue and green in Figure 4b. For such 45°-structure, the diffraction patterns both measured experimentally Figure 4e and calculated by the "points of intersections" model Figure 4h have the $C_{4v}$ symmetry in contrast to the $C_2$ diffraction patterns calculated using the "point scatterers along the rod" model, Figure 4k. Note that all results of calculations using the "points of intersections" model Figure 4g-i are in excellent agreement with experimental data presented in Figure 3 and 4d-f.

With this point clear we may go ahead to the interpretation of all experimental data. From Figure 4c one can see that in the anisotropic glassy woodpile all scatterers (points of intersections) owned by each *x*-oriented rod are distributed equidistantly (blue points in Figure 4c due to equidistant arrangement of the *y*-oriented ordered set of rods. Therefore each *x*-oriented rod gives rise to one perfect $C_{4v}$ diffraction pattern with the symmetry axis turned about its center by random angle $\alpha_i$ with respect to the ordered state $\alpha_i = 0$. That is why it possible to observe both zero- and first-order scattering from disordered set of rods for all investigated anisotropic structures, Figure 3. Now the effect of disappearance of the arcs related to the ordered *y*-oriented set of rods becomes straightforward. Because of the disorder in the *x*-oriented set of rods, all interscatterers distances in *y*-direction become randomized (red points in Figure 4c) resulting in the destruction of the perfect arcs. Small clusters of scatterers with sufficiently close interscatterer distances determine weakly arcs that can be seen in the diffraction patterns under close examination Figure 3. The random angle of the corresponding first-order scattering cone is defined by the equation $\theta_{s1} = \cos^{-1}(\lambda / a_{random})$.

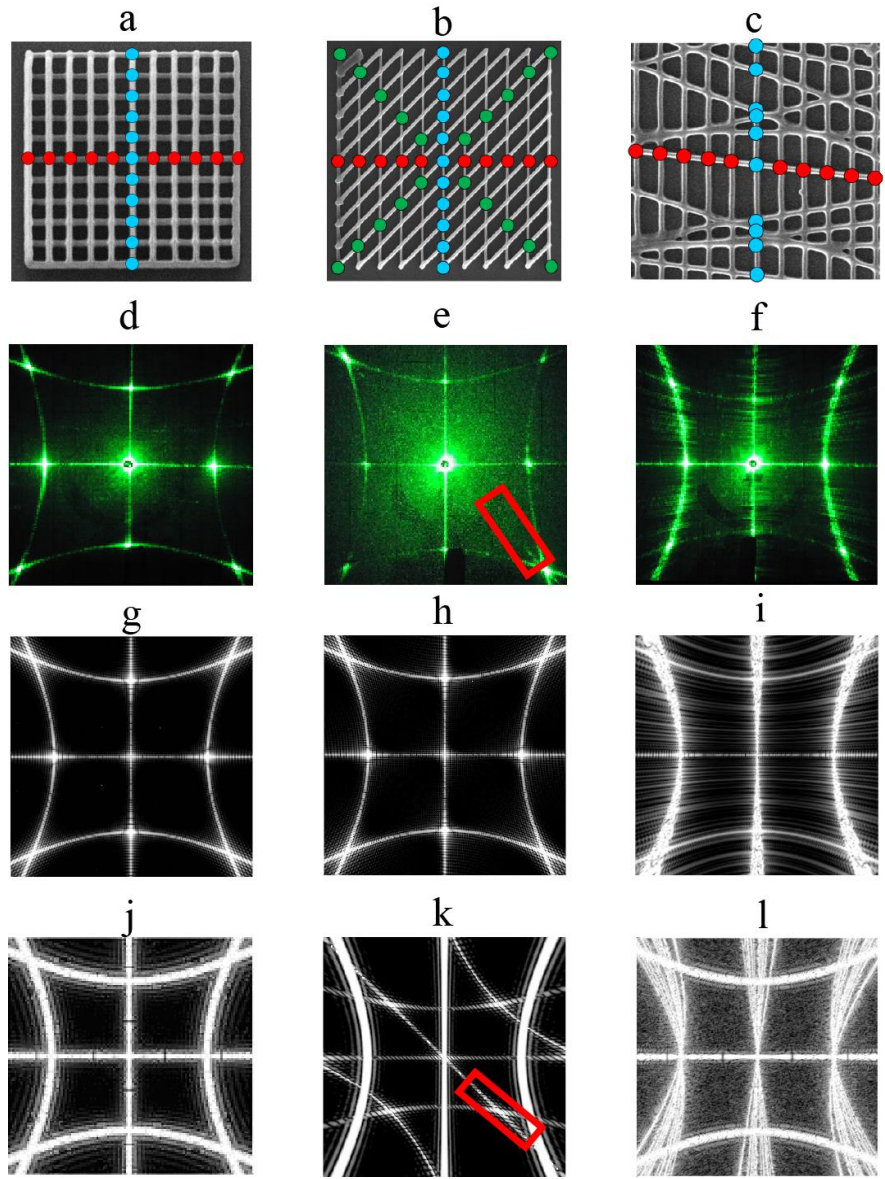

FIG. 4. Schematic of the square ordered (a), 45°-ordered (b), and disordered (c) woodpile-type photonic structures. (d-f) Experimentally measured diffraction patterns corresponding to the structures from upper series. $a_x = a_y = 1 \mu m$, $\lambda = 0.53 \mu m$. (g-i) Diffraction patterns numerically calculated using the "points of intersections" model. (j-l) Diffraction patterns numerically calculated using the "point scatterers along the rod" model for 1000 scatterers along the rod 50 µm long. Calculated diffraction patterns from third and fourth rows correspond to the structures from upper row.

Next, we consider an impact of the scattering from whole rods on the diffraction patterns of woodpile structures. To clarify this point we compare results for the 45°-ordered woodpile structure. Red rectangles mark a well detectable diagonal stripe in the pattern calculated by using the model of "point scatterers along the rod" shown in Figure 4(k) and a very weak diagonal stripe in the experimental pattern demonstrated in Figure 4(e). The comparison of panels (h) and (k) in Fig.4 demonstrates that the diagonal stripe in the diffraction pattern is the fingerprint of the scattering from rods.

## VI. DISORDER-INDUCED LIGHT SCATTERING FROM GLASSY METASURFACES

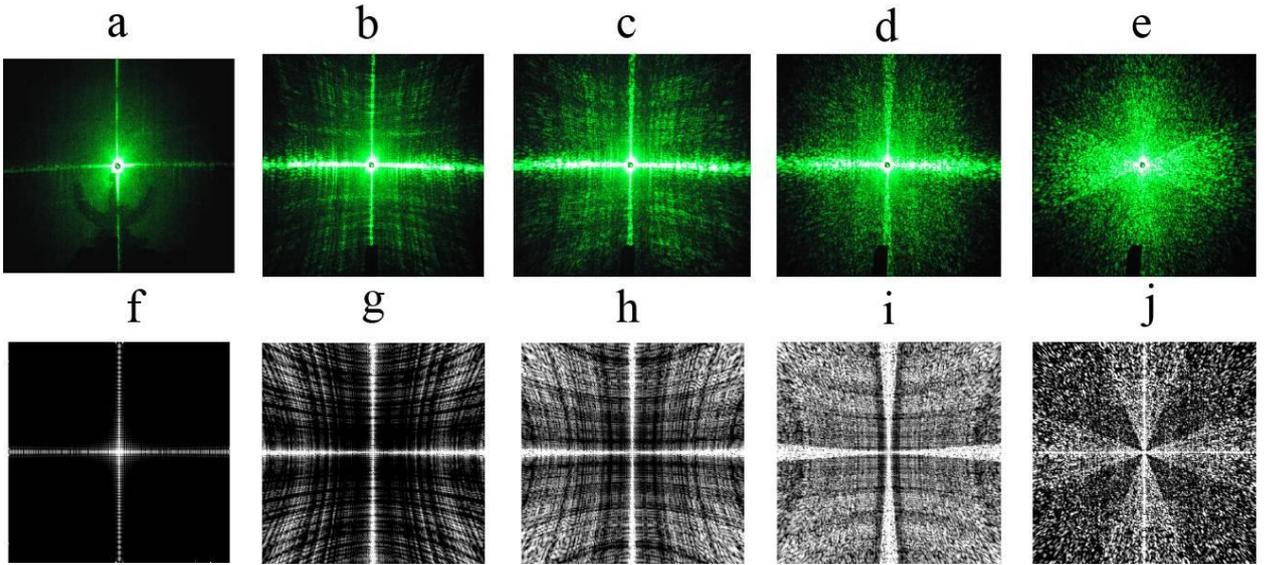

FIG. 5. Diffraction pattern evolution for glassy woodpile metasurface depending on the disordered parameter $p$ for the normal distribution. a) $p = 0$, b) $p = 0.02$, c) $p = 0.04$, d) $p = 0.1$, e) $p = 0.5$. All structures have external size in the $xy$ plane of $50 \times 50\,\mu m$, the lattice parameters of the ordered sample $a_x = a_y = 0.5\,\mu m$, the number of layers along the $z$-axis $N=2$. $\lambda = 0.53\,\mu m$. f-j) Diffraction patterns numerically calculated using the "points of intersections" model for structures from upper series.

Figure 5 shows the transformation of the diffraction patterns of the glassy woodpile metasurface for a range of levels of disorder. For ordered metasurface under condition $a_x = a_y < \lambda$ only the zero-order diffraction ($n=0$) is allowed with the angle of light scattering $\theta_s = 90°$. As a result the only perfect cross is observed on a far-field screen, Figure 5a. When we introduce the random fluctuations of the orientation for both $x$-oriented and $y$-oriented individual rods, the diffraction patterns on the screen changed dramatically. The new property is that a great number of sufficiently strong arcs appear on the screen along both x- and y-axis exactly as forbidden in perfect structure first-order scattering, Figure 5b-c. The reasons for this effect can be understood in terms of the previous discussion. As a result of the disorder, a number of interscatterers distances both in *x*- and *y*-direction have become larger than laser wavelength $\lambda < a_{random}$. Therefore the necessary condition for first-order diffraction $\lambda < \alpha < 2\lambda$ is fulfilled and disorder-induced arcs with the random angles $\theta_{s1} = \cos^{-1}(\lambda/a_{random})$ are observed. For larger levels of disorder the speckle-like pattern extends over the whole screen, Figure 5d-e.

A careful analysis of the diffraction patterns shown in Figures 3 and 5 reveals the intensity enhancement of the high-order patterns with increasing of the disordered parameter *p*. This effect can be explained by taking into account that the square *S* of the intersection region of two rods depends strongly on the intersection angle α. For the perfect woodpile structure α=90° and $S=w^2$ where *w* is the rod width. For the disordered structure $S = w^2/\sin\alpha$ and the larger parameter *p* corresponds to the larger deviation from α=90° that leads to the increasing of the intersection area *S* and the patterns become more and more intensive.

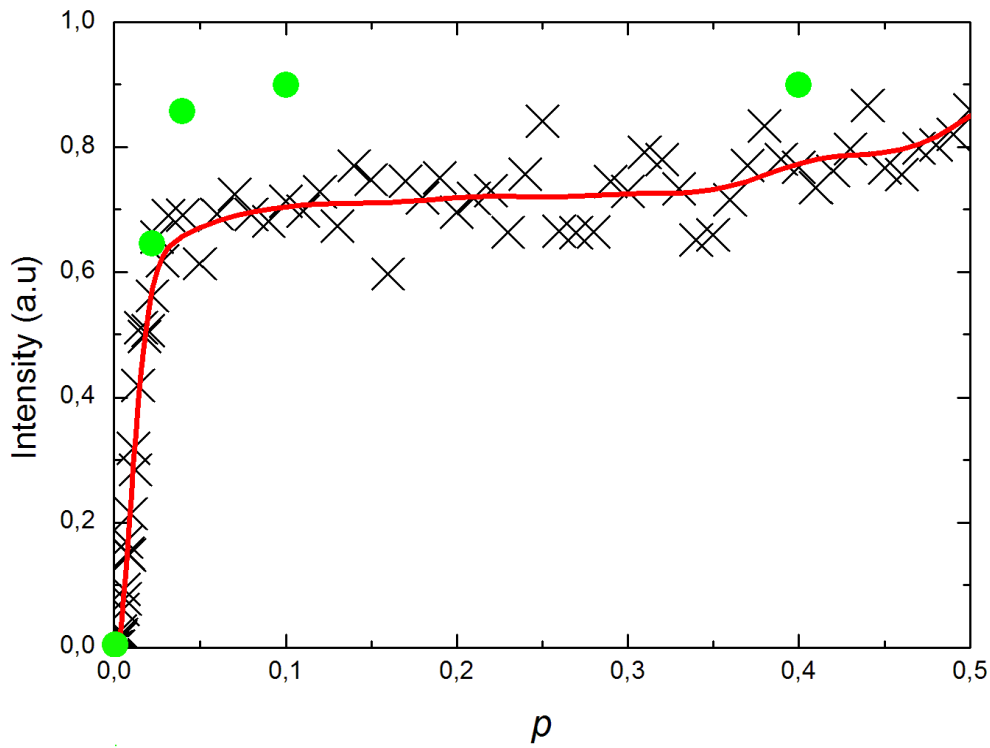

FIG. 6. Calculated and experimental intensity of the total disorder-induced scattering depending on the disordered parameter $p$ for metasurfaces with the normal distribution (the standard deviation is $\sigma = p \cdot \pi/4$). All structures have external size in the $xy$ plane of $50 \times 50 \mu m$, the lattice parameters of the ordered sample $a_x = a_y = 0.5 \mu m$, the number of layers along the $z$-axis $N=2$; $\lambda = 0.53 \mu m$. The black crosses are scattering intensity values, the red curve is a guide for eyes only, and the green circles are normalized experimental data.

Figure 6 presents the experimental and simulated intensity of the total disorder-induced first-order scattering depending on the disordered parameter $p$ for metasurfaces with the normal distribution function (the standard deviation is $\sigma = p \cdot \pi/4$). This dependence shows extremely strong increase of the scattering intensity for a small degree of disorder ($0 < p < 0.05$). Therefore, we obtain a simple and sensitive test of perfection for woodpile structures that is important for variety of applications. It just requires that one should choose the laser wavelength $\lambda$ a bit longer than the lattice constant $a$. Otherwise the experiment satisfies the condition of a deep metasurface regime $\lambda \gg a$ and the traces of arcs appear only at a high disorder degree values. This test can be employed for functional woodpile structures in biomedicine and biology[27, 28], design of filtration membranes for nano- and micro- level separation, a planar antennas[29] etc.

## VII. CONCLUSION AND OUTLOOK

We demonstrate that diffraction experiments offer unique information on the mechanisms of light scattering from ordered and disordered dielectric photonic structures. The measurements were carried out on thin photonic slab and two-layered woodpile-based metasurfaces, the results were discussed on the basis of the Laue equations in the two-dimensional approximation. The experimental results allow us to interpret the diffraction patterns and underline the specific features arising as a result of elegant interplay between order and disorder.

In photonic structures, the zero-order and high-order diffraction patterns show different behavior as a function of the disorder parameter. The zero-order diffraction ($n=0$) is not significantly modified for small and intermediate disorder, but the higher orders of diffraction ($n= \pm1, \pm2, \pm3...$) are strongly affected even for small disorder. In addition, for anisotropic structures when orientational disorder is introduced only in one direction of square woodpile, we found unexpected effect. When amount of disorder increases, the high-order diffraction patterns from ordered set of rods becomes more and more randomized and finally hardly observed while high-order patterns from disordered set of rods continue to be bright and sharp nearly independent on the level of disorder. We demonstrate that the reason for such effect is that the light scattering can be described purely in terms of the intersection points of the rods. This conclusion can be considered as general features of light scattering in dielectric photonic structures and lead to important new understanding about light propagation in random photonic medium.

Moreover, these investigations are of interest because strongly disordered systems offer much optical functionality; possess unexpected physical properties, which make them potentially useful as an alternative to the pure periodic structures. We can conclude that in a photonic structure the scattering intensity is defined by the number of intersection points rather than by the structural building elements as a whole. It is sufficient to mention such application as random lasing from photonic glasses[30], laser speckle techniques for a variety of optical metrology techniques[15] and a possible new generation of solar cells basing on the interaction of ordered and disordered elements that are much more efficient at trapping and absorbing sunlight[31].

ACKNOWLEDGEMENTS

We thank S.Y. Lukashenko for the technical help. This work was supported by the Russian Foundation for Basic Research (Grant No. 18-02-00427), by the Ministry of Education and Science of Russian Federation (Project No. 3.1500.2017/4.6), and by Presidium of RAS (Program 32, Nanomaterials).